\documentclass{article}
\usepackage{spconf,amsmath,graphicx}

\usepackage{enumitem}
\setlist{nosep, leftmargin=14pt}

\usepackage{mwe} 
\usepackage{algorithm}
\usepackage{algpseudocode}
\usepackage[table]{xcolor}
\usepackage{subcaption}
\usepackage{graphicx}
\usepackage{import}
\usepackage{svg}
\usepackage{siunitx}
\usepackage{hyperref}
\usepackage{fancyhdr}

\newcommand{\redbox}[1]{\fcolorbox{red}{white}{$#1$}}
\newcommand{\bluebox}[1]{\fcolorbox{blue}{white}{$#1$}}

\title{Reconstructing Patched or Partial Holograms to allow for Whole Slide Imaging with a Self-Referencing Holographic Microscope}
\name{Philip Groult$^\ast$, Julia D. Sistermanns$^\ast$, Ellen Emken\textsuperscript{\textdagger}, Oliver Hayden\textsuperscript{\textdagger}, Wolfgang Utschick$^\ast$}
\address{$^\ast$\normalsize{Chair of Methods of Signal Processing, TUM School of CIT, Technical University of Munich, Germany} \\ \textsuperscript{\textdagger}\normalsize{Heinz-Nixdorf-Chair of Biomedical Electronics, TUM School of CIT, Technical University of Munich, Germany}\\\small{Email Corresponding Author: julia.sistermanns@tum.de}}
\begin{document}
%
\maketitle
\fancypagestyle{cfooter}{ %
\fancyhf{} 
\cfoot{\scriptsize{© 2026 IEEE. Personal use of this material is permitted.  Permission from IEEE must be obtained for all other uses, in any current or future media, including reprinting/republishing this material for advertising or promotional purposes, creating new collective works, for resale or redistribution to servers or lists, or reuse of any copyrighted component of this work in other works.}
}
\renewcommand{\headrulewidth}{0pt} 
\renewcommand{\footrulewidth}{0pt}
}
\thispagestyle{cfooter}
\begin{abstract}
The last decade has seen significant advances in computer-aided diagnostics for cytological screening, mainly through the improvement and integration of scanning techniques such as whole slide imaging (WSI) and the combination with deep learning. Simultaneously, new imaging techniques such as quantitative phase imaging (QPI) are being developed to capture richer cell information with less sample preparation. So far, the two worlds of WSI and QPI have not been combined. In this work, we present a reconstruction algorithm which makes whole slide imaging of cervical smears possible by using a self-referencing three-wave digital holographic microscope. Since a WSI is constructed by combining multiple patches, the algorithm is adaptive and can be used on partial holograms and patched holograms. We present the algorithm for a single shot hologram, the adaptations to make it flexible to various inputs and show that the algorithm performs well for the tested epithelial cells.
\end{abstract}
\section{Introduction}
\label{sec:intro}
\begin{figure}[t]
    \centering
    \includegraphics[height=3cm]{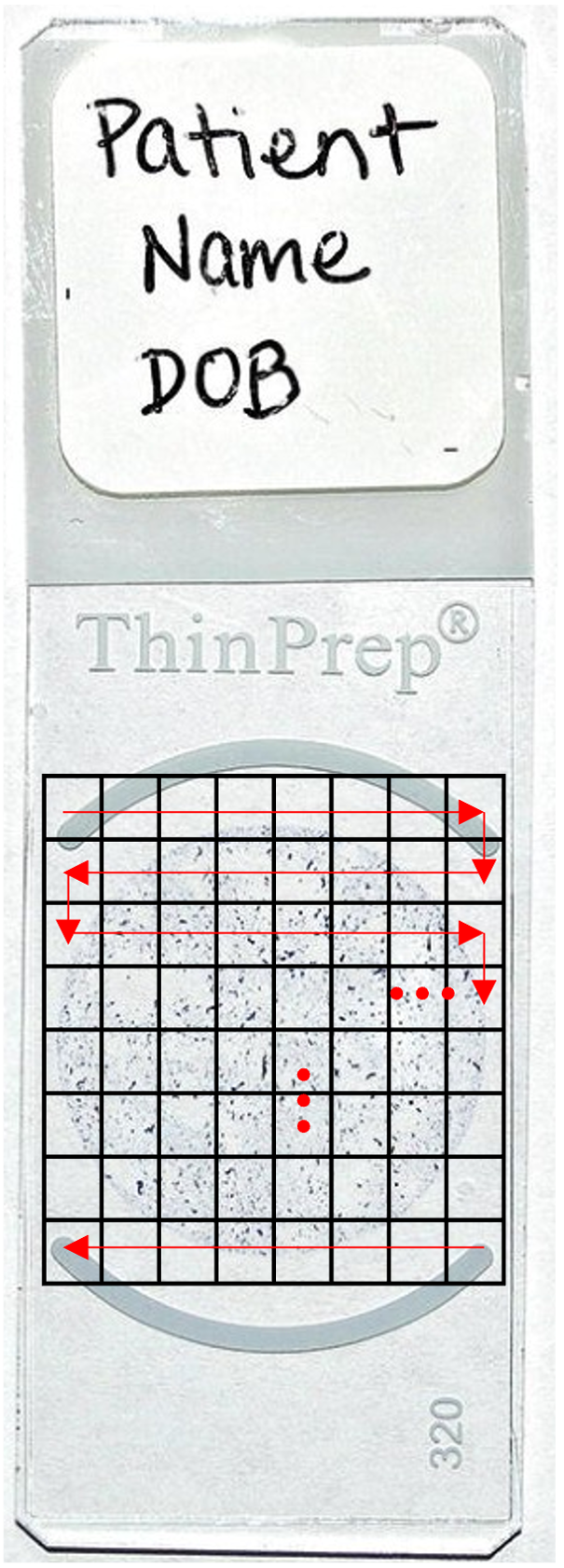}
    \includegraphics[height=3cm]{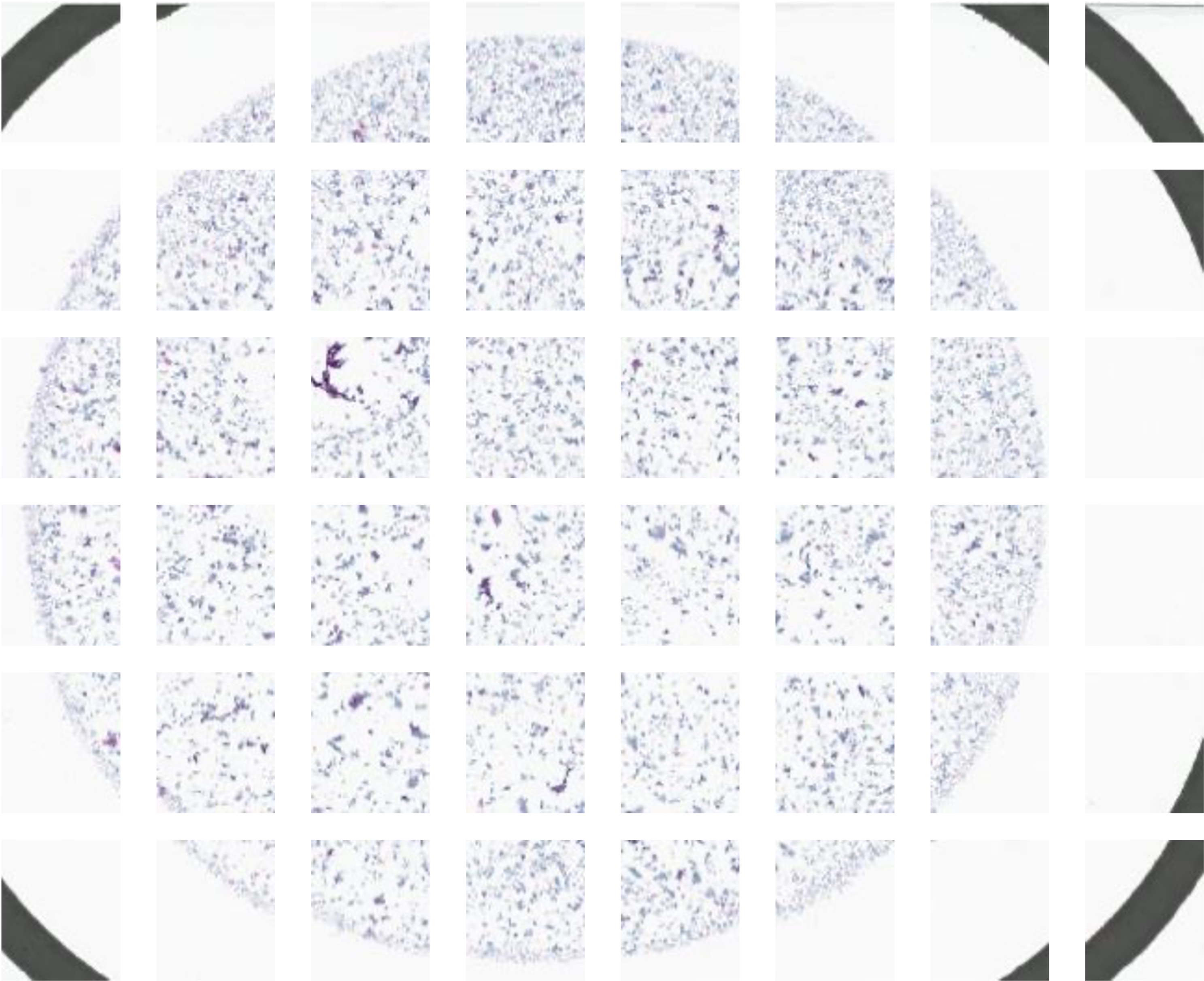}
    \hspace{6mm}
    \includegraphics[height=3cm]{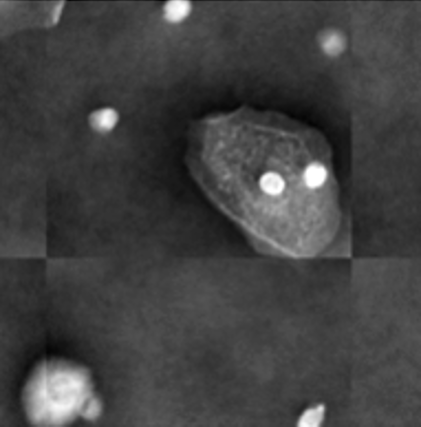}
    \caption{Left: Visualization patch-wise WSI, Right: Example cell for patch-wise reconstruction}
    \label{fig:patch-mic}
    \vspace{-0.5cm}
\end{figure}

Computer-aided diagnostics (CAD) for cytological screening have seen great advances in the past few years, integrating Deep Learning (DL) techniques into the diagnostic process~\cite{Al-Janabi:2012aa, DL}. CAD techniques require sufficient amounts of high-quality data~\cite{Al-Janabi:2012aa}, which is why they rely heavily on high-throughput measurement schemes. One interesting imaging technique providing information-rich cell measurements with little sample preparation is Quantitative Phase Imaging (QPI)~\cite{QPI, roadmap}.
In QPI, the complete object wave can be recorded, providing both the amplitude and phase information of a cell. It is label-free and non-destructive, allowing for live cell imaging \cite{live_cell}. QPI methods, however, require object wave reconstruction. The currently available reconstruction algorithms for QPI generally focus on reconstructing single-shot holograms \cite{off_axis, Schnars, reconstruction, roadmap, convex_it, GNN}. The reconstruction depends on the configuration of the optical path. 
While QPI works well for single-shot recording of fixed-size holograms \cite{QPI, Klenk:2023ab}, the current standard of digitized samples for CAD techniques is Whole Slide Imaging (WSI). In WSI a slide scanner is used to scan a stained and fixed sample using a brightfield microscope, providing an image familiar to a cytopathologist~\cite{Al-Janabi:2012aa}. The slide is recorded digitally by patching multiple smaller images in a grid-like pattern \cite{WSI} (Fig.~\ref{fig:patch-mic}). To establish QPI as an alternative high-throughput measurement technique for CAD a comparison to the gold standard on a cell-by-cell basis is required. But WSI poses several new challenges for QPI as the reconstruction is especially sensitive to objects and artifacts at the edges of a recorded hologram~\cite{ILS, example_SF3}. Naturally, this occurs frequently in a WSI situation as the patches are fixed and not adapted to the sample, leading to discontinuities at almost all patch lines. For cells or cell clusters at these patch lines, this results in odd internal structures and implausible phase values within the cell (Fig.~\ref{fig:patch-mic}). Consequently, single-cell analysis is limited due to the low data quality.

In this work, whole slide images were recorded with a self-referencing three-wave digital holographic microscope (DHM). Our contribution is an algorithm to reconstruct the resulting holograms. It is adapted to allow for holograms of arbitrary size: cutouts containing a single cell or cluster and cutouts of patched holograms. It will be shown that single-cell reconstruction of cutouts from holograms is possible while maintaining the quality of the reconstruction. 
Several existing strategies were tested to approach and overcome the new types of artifacts and noise resulting from the patching, and an adaptation to the phase integration was developed to reach a high reconstruction quality of a whole slide hologram.

\section{Data}
\label{sec:data}
The data used to develop and test the algorithm was taken from an ongoing clinical study conducted by the Chair of Methods of Signal Processing in cooperation with the Institute for Pathology, and the Heinz-Nixdorf Chair for Biomedical Electronics at the Technical University of Munich.

The study uses a self-referencing three-wave digital holographic microscope with lateral shear~\cite{Emken:2025aa}. All holograms are recorded as a single-shot and have a fixed dimension of $2048\times 1536$ pixels. 
To make WSI possible using such a setup, a standard stained and fixed sample slide of a cervical smear is placed under the microscope and the object carrier is moved automatically, recording a single shot at each position. After the recording, the holograms are patched together to get a patched hologram of the full slide (see Fig.~\ref{fig:patch-mic}).

QPI is able to provide richer cell information than a simple grayscale brightfield image, as not just the amplitude, but the phase information of the object is recorded. The reconstructed phase $\phi$ is of special interest because it is tied to actual physical quantities: the local thickness of the cell and the difference in refractive indexes between the cell and the surrounding medium~\cite{Girshovitz}. We refer to the combination of these physical quantities as the so-called \textit{optical height} $h_{opt}$, sometimes referred to as the optical path delay~\cite{Girshovitz}:
\begin{align}
     h_{opt}= \phi \cdot \frac{\lambda}{2\pi}.
\end{align}
$h_{opt}$ is given in $\mu \text{m}$ and is computed using the wavelength $\lambda$, which in our setup is $\lambda = \SI{528}{\nano\m}$.
\begin{figure}[t]
    \begin{subfigure}[c]{0.4\linewidth}
    \includegraphics[width=\linewidth]{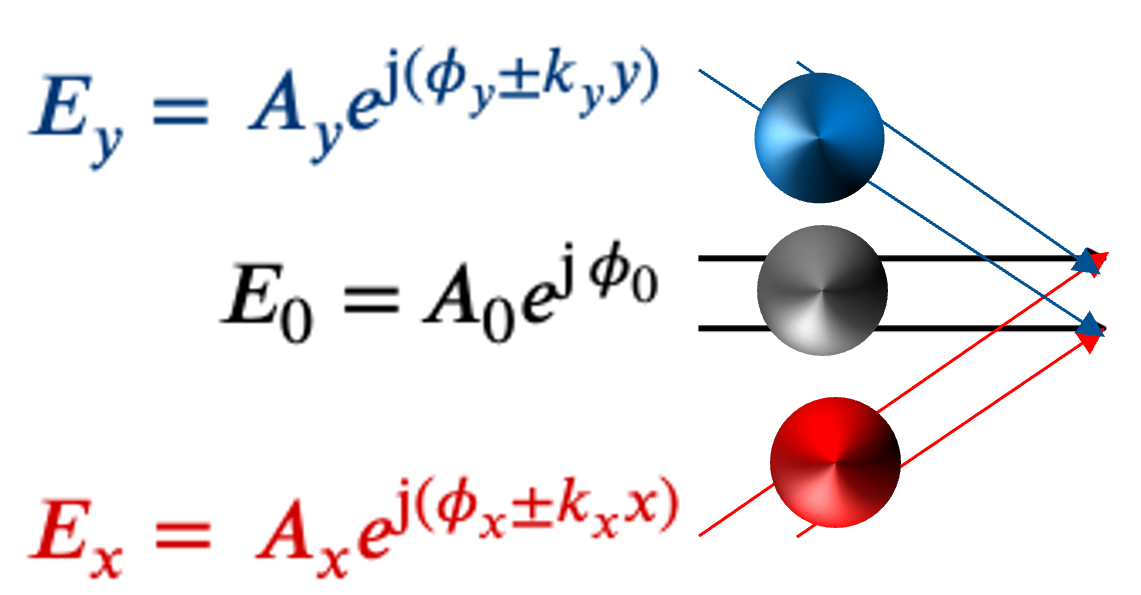}
\end{subfigure}%
\hfill
    \begin{subfigure}[c]{0.55\linewidth}
    \includegraphics[width=\linewidth]{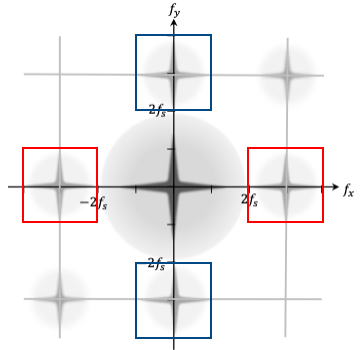}
\end{subfigure}
    \caption{Beams of the self-referencing three-wave lateral shear DHM and corresponding Fourier spectrum}
    \label{fig:self-ref-spectrum}
    \vspace{-0.3cm}
\end{figure}

\section{Methodology}
\label{sec:methodology}

In our self-referencing three-beam setup, the hologram records the interaction between the unshifted object beam $E_O$ and two shifted object beams ($E_y$ and $E_x$) (Fig.~\ref{fig:self-ref-spectrum}). The recorded interference pattern can be expressed using the following eq.~[\cite{Sistermanns:2025aa} (3.66-3.67)]:
\begin{align}
I(x,y) = &|E_O+E_x+E_y|^2, \\
I(x, y) = & A_O^2+A_x^2+A_y^2 \nonumber\\
& +A_y A_x (e^{j(\phi_x-\phi_y)} +e^{-j(\phi_x-\phi_y)}) e^{\pm j(k_xx+k_yy)} \nonumber\\
&+\redbox{A_O A_x (e^{j(\phi_x-\phi_O)} +e^{-j(\phi_x-\phi_O)}) e^{\pm j k_xx}} \nonumber\\
&+\bluebox{A_O A_y (e^{j(\phi_y-\phi_O)} +e^{-j(\phi_y-\phi_O)}) e^{\pm j k_yy}}.
\label{gradients}
\end{align}

\subsection{Gradient extraction and calibration}
The terms of Eq.~\ref{gradients} correspond to separate lobes in the Fourier Spectrum for well-chosen $k_x$ and $k_y$ (Fig.~\ref{fig:self-ref-spectrum}). To reconstruct the amplitude information, the central lobe is extracted, which corresponds to the sum of the intensities of the three beams. The quantitative phase information can be reconstructed by extracting and then integrating the phase gradients from the spectrum~\cite{Sistermanns:2025aa}. 

The gradient in $x$-direction $\frac{\partial}{\partial x} W$ and its complex conjugate correspond to the red lobes in the spectrum in Fig.~\ref{fig:self-ref-spectrum} (also red in Eq.~\ref{gradients}). Similarly, the gradient in the $y$-direction $\frac{\partial}{\partial y} W$ is shown in blue. To extract these lobes, their center point needs to be found. For the standard, fixed-size hologram, the positions of the lobes are constant.
When inputting holograms of variable size, the position and radii of the lobes change. To adapt the algorithm, we exploit that the maximum power of a lobe is at its center. So the algorithm searches for the maximum power in the corresponding area in the spectrum. The extraction is performed with a square window of a variable size $R$. For a hologram of dimension $N \times M$, it is chosen as $R = 0.1 \min (N,M)$. The gradients are calibrated to account for static noise by using an object-free recording~\cite{Sistermanns:2025aa}. For a whole slide hologram, this calibration image needs to be adapted to match the patch in size and patch lines.
 
\subsection{Phase integration}


As mentioned, phase integration is needed to get the quantitative phase information from the obtained gradients. Two main approaches exist: the Iterative-Least Square Method of Integration (ILS) and the Complex Plane Method of Integration~\cite{ILS}. ILS was chosen as it was found to be the more robust method. 
With a change in the hologram dimensions, the frequency coordinates $f_x$ and $f_y$ change as they are directly proportional. The ILS can be written as~[\cite{Sistermanns:2025aa} (3.74)]:
\begin{align}
     W = \mathcal{F}^{-1}\left\{ \frac{f_x*\mathcal{F}\left\{\frac{\partial}{\partial x} W \right\}+f_y*\mathcal{F}\left\{ \frac{\partial}{\partial y} W \right\}}{f_x^2+f_y^2} \right\}
     \label{ils}.
\end{align}
\begin{figure}[t]
\textcolor{black}{
    \begin{subfigure}[b]{0.35\linewidth}
        \def\svgwidth{2\linewidth}
        \resizebox{!}{4cm}{
\begingroup%
  \makeatletter%
  \providecommand\color[2][]{%
    \errmessage{(Inkscape) Color is used for the text in Inkscape, but the package 'color.sty' is not loaded}%
    \renewcommand\color[2][]{}%
  }%
  \providecommand\transparent[1]{%
    \errmessage{(Inkscape) Transparency is used (non-zero) for the text in Inkscape, but the package 'transparent.sty' is not loaded}%
    \renewcommand\transparent[1]{}%
  }%
  \providecommand\rotatebox[2]{#2}%
  \newcommand*\fsize{\dimexpr\f@size pt\relax}%
  \newcommand*\lineheight[1]{\fontsize{\fsize}{#1\fsize}\selectfont}%
  \ifx\svgwidth\undefined%
    \setlength{\unitlength}{962.19498762bp}%
    \ifx\svgscale\undefined%
      \relax%
    \else%
      \setlength{\unitlength}{\unitlength * \real{\svgscale}}%
    \fi%
  \else%
    \setlength{\unitlength}{\svgwidth}%
  \fi%
  \global\let\svgwidth\undefined%
  \global\let\svgscale\undefined%
  \makeatother%
  \begin{picture}(1,0.87858267)%
    \lineheight{1}%
    \setlength\tabcolsep{0pt}%
    \put(0,0){\includegraphics[width=\unitlength,page=1]{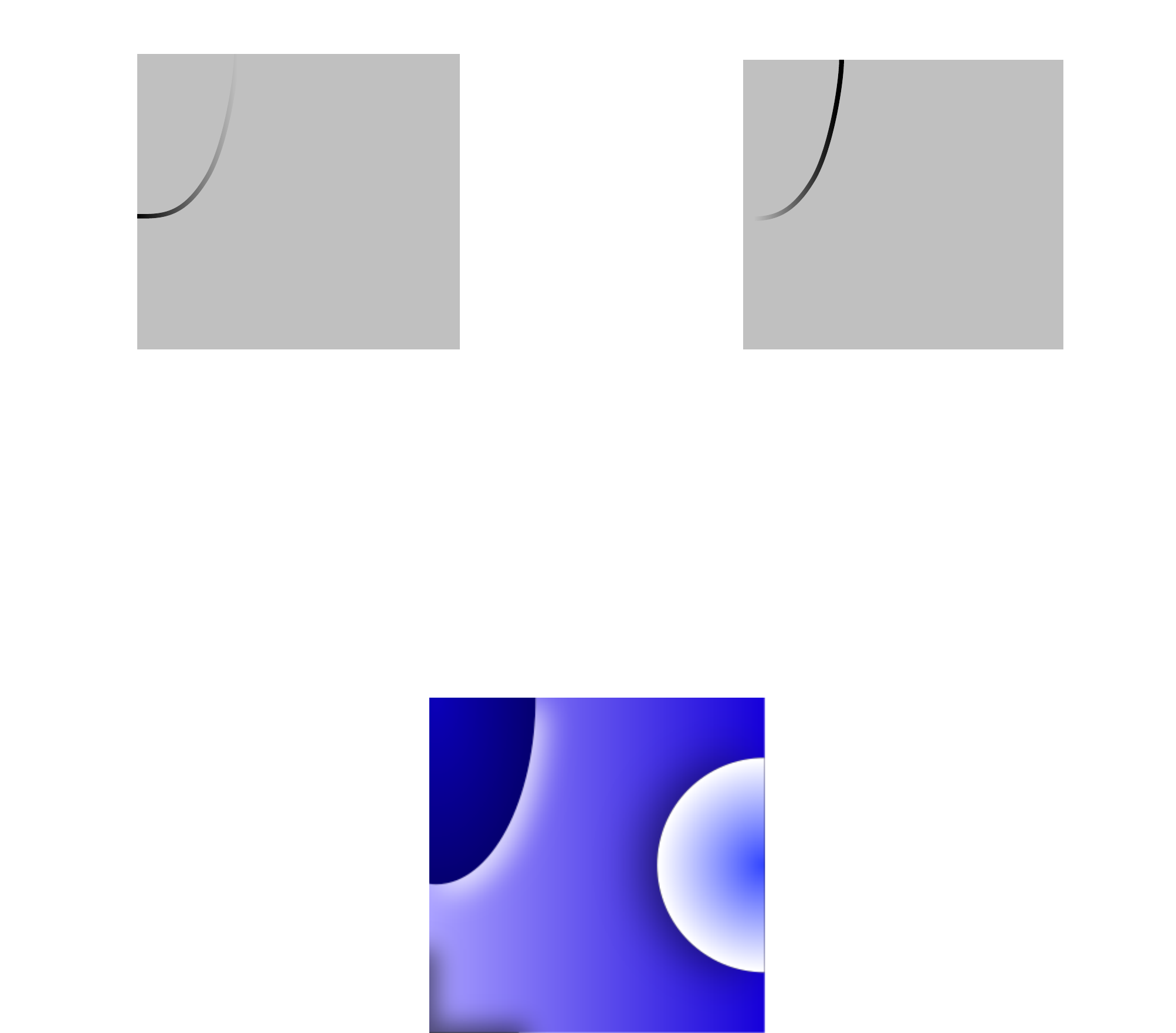}}%
    \put(0.18408263,0.84716932){\color[rgb]{0,0,0}\makebox(0,0)[lt]{\lineheight{1.25}\smash{\begin{tabular}[t]{l}$\partial_x W$\end{tabular}}}}%
    \put(0.6967133,0.84572512){\color[rgb]{0,0,0}\makebox(0,0)[lt]{\lineheight{1.25}\smash{\begin{tabular}[t]{l}$\partial_y W$\end{tabular}}}}%
    \put(0.1000348,0.14117615){\color[rgb]{0,0,0}\makebox(0,0)[lt]{\lineheight{1.25}\smash{\begin{tabular}[t]{l}$W-W_h$\end{tabular}}}}%
    \put(0,0){\includegraphics[width=\unitlength,page=2]{phase_integration_basic_isbi.pdf}}%
    \put(0.33064317,0.44142605){\color[rgb]{0,0,0}\makebox(0,0)[lt]{\lineheight{1.25}\smash{\begin{tabular}[t]{l}Phase Integration\end{tabular}}}}%
    \put(0,0){\includegraphics[width=\unitlength,page=3]{phase_integration_basic_isbi.pdf}}%
  \end{picture}%
\endgroup%
}
    \end{subfigure}
    \hfill
    \begin{subfigure}[b]{0.45\linewidth}
        \def\svgwidth{1.3\linewidth}
        \resizebox{!}{4cm}{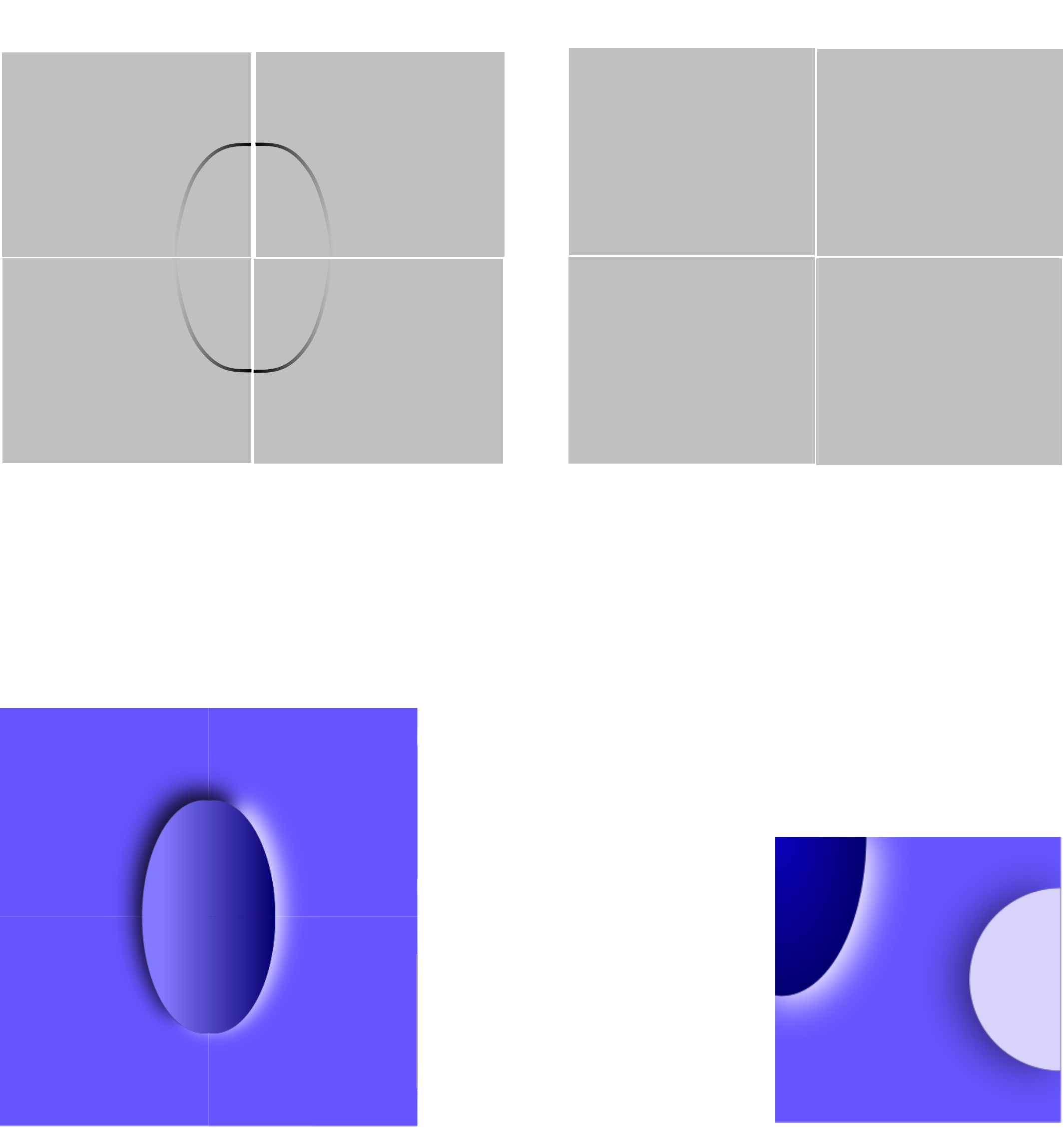}
    \end{subfigure}}
    
    \caption{Phase integration without and with accounting for boundary artifacts (adapted from~\cite{ILS,Sistermanns:2025aa})}
    \label{fig:extended}
\end{figure}
Phase integration relies on the assumption of a continuous wavefront differentiable over the entire frequency domain~\cite{integration2}. Any discontinuities within the hologram, typically found at the edges, lead to significant noise. To remove edge discontinuities, the gradients can be extended by replicating and mirroring to form a larger continuous spectrum with either Mirrored Derivative Integration (MDI) (see Fig.~\ref{fig:extended}) or Antisymmetric Derivative Integration~\cite{ILS}. Out of the two, MDI showed better results and a lower computational complexity, so it was chosen for this work. 

\subsection{Method of evaluation}
\label{evaluation}
Three different types of holograms were analyzed: standard-size holograms serving as a reference, cutouts of cells, and patched holograms. For the cutouts, the phase is reconstructed with and without MDI and is compared to the corresponding area containing the cell in the original hologram. A mask of the cell is applied to remove the background from the comparison. Two different measures are computed: the normalized $L_1$-norm of the local errors between the phase values of the cutout $p_c$ and the standard hologram $p_0$ and the mean difference $\varepsilon_\mu$:
\begin{align}
     & L_1 = \frac{1}{N_p}\sum |p_{\text{0}}-p_{\text{c}}| \\
     & \varepsilon_\mu = \frac{1}{N_p}\sum |p_{\text{0}}| - \frac{1}{N_p}\sum |p_{\text{c}}|.
    \label{diff_measurement}
\end{align}
The $L_1$-norm was utilized as the obtained values are smaller than zero and the $L_2$-norm would reduce the total error~\cite{norm}.


As no ground truth or reference exists for the patched holograms, the artifacts that occurred were counted. The focus was on cells on the patch lines, as this affects the clarity of inner structures. Additionally, the optical height range was evaluated to ensure realistic values for the given cell types.


\section{Results}
\label{sec:Results}
\begin{figure}[t]
    \centering
    \includegraphics[width=1\linewidth]{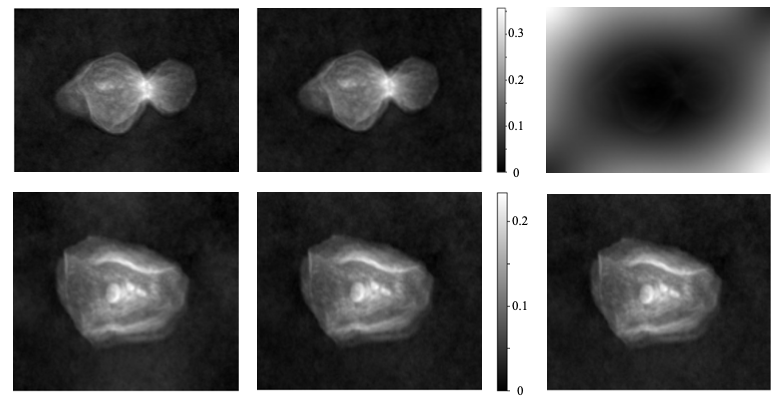}
    \caption{Reconstruction phase for cutout with MDI (left), for the standard size (middle) and without MDI (right)}
    \label{fig:cutouts}
\end{figure}

\bgroup
\def\arraystretch{1.3}
\begin{table}[t]
\centering
\begin{tabular}{l|c|c|c|c}
\rowcolor[HTML]{C0C0C0}
 & $L_1$ & $\varepsilon_{\mu}$ & $L_{1,\text{MDI}}$ & $\varepsilon_{\mu, \text{MDI}}$ \\ \hline
Mean & 0.98723 & 0.98561 & 0.00927 & 0,00872 \\ \hline
Max  & 4.27265 & 4.27265 & 0.07535 & 0.07535 \\ \hline
Min & 0.00140 & 0.00004 & 0.00077 & 0.00000 \\ \hline
Var & 1.07981 & 1.08216 & 0.00012 & 0.00013 \\ \hline
Median  & 0.79002 & 0.79002 & 0.00558 & 0.00533 \\ 
\end{tabular}
\caption{Pixelwise error ($L_1$) and mean difference ($\varepsilon_{\mu}$) between original and cutouts with and without MDI in $\SI{}{\micro\m}$}
\label{tab: comparison}
\vspace{-0.3cm}
\end{table}
\egroup

\begin{figure*}[t]
    \centering
    \includegraphics[width=1\textwidth]{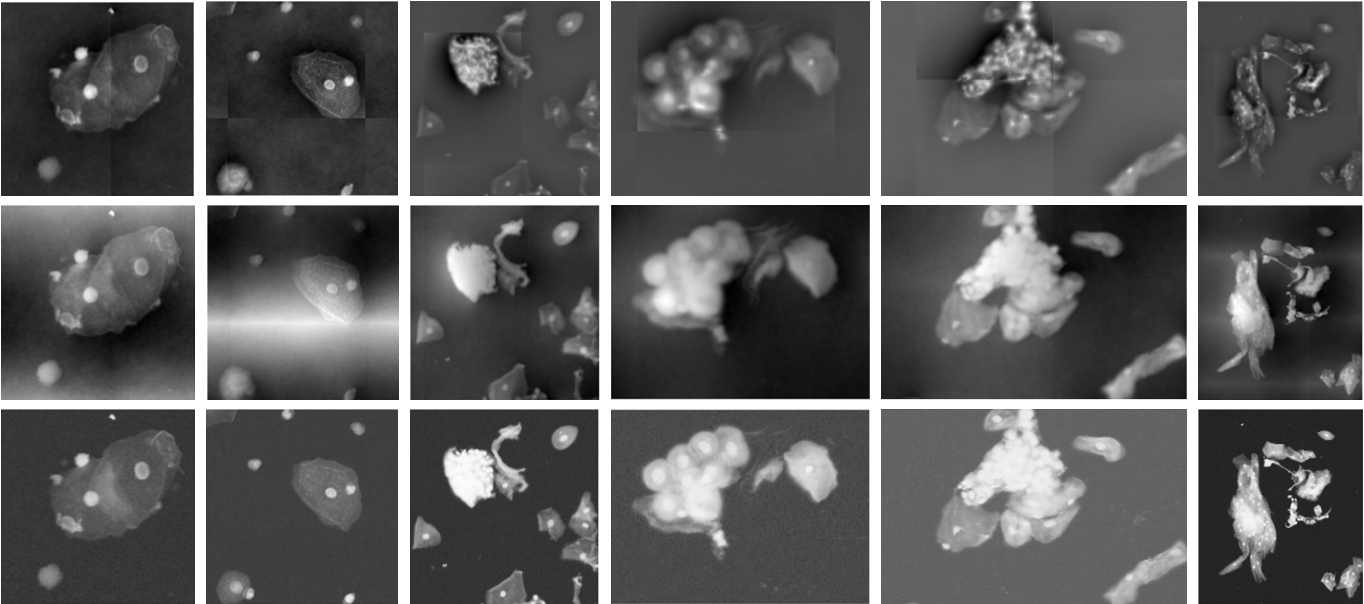}
        \caption{Row 1: (1.) patching after reconstruction, Row 2: (2.) reconstructing patched hologram as one using algorithm with MDI, Row 3: (3.) reconstructing patched hologram as one with adapted phase integration}
            \label{fig:patch}
\end{figure*}

In Sec.~\ref{sec:results-single}, standard sized holograms were compared to cutouts. For this purpose, 125 cells or cell clusters recorded using the DHM and a microfluidic channel were chosen. A typical size of a cutout is around $600\times 400$ pixels ($\approx 0.06$ times the size of the standard hologram). 

In Sec.~\ref{sec:results-wsi}, 20 large patches from multiple WSI's acquired with the DHM were reconstructed, and 169 cells or cell clusters on patch lines were evaluated. The patches have various sizes; the largest is $ 1.69 \ \text{GB}$.\footnote{Souce code available at \url{https://github.com/juliaSis/reconstruct-patched-or-partial-wsi-holograms}.}
\vspace{-0.2cm}
\normalsize
\subsection{Cutouts of standard size holograms}
\label{sec:results-single}
Table \ref{tab: comparison} shows the two error metrics defined in Sec.~\ref{evaluation} when comparing the cell cuts to the full reconstruction for the simple phase integration and using MDI. The average optical height range for the analyzed epithelial cells is $[0, \SI{0.25}{\micro\m}]$, with a maximum observed value of $\SI{0.65}{\micro\m}$. 

Looking at the values in Table \ref{tab: comparison}, it is apparent MDI greatly reduces the average error for most cells. A relation between the two error metrics is also apparent, as they are nearly equal for most cells. Performing the reconstruction of the cutout using MDI consistently results in a high-quality reconstruction. Fig.~\ref{fig:cutouts} shows the cells reconstructed from cutouts with MDI are visually indistinguishable to the cutout from the full reconstruction, while large errors ($\leq \SI{4}{\micro\m}$) can occur when performing the simple integration (example on the upper right). The primary drawback of using MDI is the fourfold increase in computational power. For some cells, this increase is not necessary as the cells are accurately reconstructed without MDI (Fig.~\ref{fig:cutouts} bottom right). This was the case for $20 \%$ of the analyzed cells. It should be noted that using a margin around the cell boundary increases robustness.


\subsection{Whole slide holograms}
\label{sec:results-wsi}
For the whole slide holograms three versions of the reconstruction are evaluated: (1.) reconstructing every single-shot hologram separately and then patching the reconstructed phase images, (2.) reconstructing the entire patched hologram as one using MDI and (3.) reconstructing the entire patched hologram as one with an adapted phase integration. The results for 6 different patches can be seen in Fig.~\ref{fig:patch}.

(1.) can result in significant artifacts at the patch lines. Especially for cells or clusters spread between multiple patches, this leads to halos, discontinuities and inconsistencies (e.g. phase values of $>\SI{1}{\micro\m}$), even in a single cell. In the majority of cases, when reconstructing the patched hologram as a whole (2.) and (3.), the phase information is more coherent, the patch lines are removed, and the inner structures of a cell are clearly visible. Although it has a slight blurring effect.

When using the algorithm as presented before for the cutouts using MDI (2.), the patch lines can still result in significant noise and reflections (e.g. Fig.~\ref{fig:patch}, row 2, image 2). It was noted that this effect is stronger with smaller patches and therefore inversely proportional to the hologram size. This is due to the assumption of a continuous wavefront and the fact that the patch lines are discontinuities. MDI only removes discontinuities at the edges, not in the middle of a patched image.
To solve this issue, an adapted method of integration (3.) was found to yield more robust results for the analyzed cells as shown in Fig.~\ref{fig:patch}. In (3.), the normalized frequency coordinates $f_x$ and $f_y$ are no longer centered on zero during the integration, but shifted off-center. 

To quantify how well (2.) and (3.) work to eliminate the artifacts at patch lines, $169$ cells and cell clusters containing artifacts in the WSI patches reconstructed with (1.) were chosen. Using method (2.) improved the reconstruction for 157 of the 169 cells, (3.) removed all patch line related issues. The typical optical height range for (2.) is $[0, \SI{0.5}{\micro\m}]$ and $[0, \SI{0.3}{\micro\m}]$ for (3.), which is slightly higher than for the non-stained cells of Sec.~\ref{sec:results-single}, but still realistic.


\section{Conclusion}
\label{sec:Conclusion}
In this work, the combination of quantitative phase and whole slide imaging was achieved using a self-referencing digital holographic microscope. To this end, an algorithm was developed to reconstruct the phase information of cells from cutouts and patched holograms directly, instead of just from the single-shot, fixed-size hologram recorded by the DHM. It was shown that using MDI during the phase integration increases accuracy and robustness, which is worthwhile, despite the increase in computational cost. For whole slide holograms, it is possible to either reconstruct the holograms one by one and patch the phase images afterwards or to patch the holograms first and reconstruct them as a whole. The main limitation is found in the patch lines, which introduce discontinuities or create artifacts in the patched phase image. When reconstructing as a whole, the effect of artifacts in cells on the patch lines is reduced. However, for some cases with strong patch lines in the middle, this leads to significant noise. To solve this, an alternative phase integration approach was introduced, which successfully removed the noise. Further work could focus on developing algorithms that eliminate such discontinuities during the patching of holograms. 

\newpage
\section{Compliance with Ethical Standards}
This study was performed in line with the principles of the Declaration of Helsinki. Approval was granted by the Ethics Committee of the Technical University of Munich (Date 25.04.2022/No. 2022-156-S-KH).

\section{Conflicts of Interest}
No funding was received for conducting this study. The authors have no relevant financial or non-financial interests to disclose.

\bibliographystyle{IEEEbib}
\bibliography{strings,refs}

\end{document}